\documentclass[referee]{aa} 
\usepackage{graphicx}

\begin{document}

\title{Magnetically coupled atmosphere, fast sausage MHD waves, and forced magnetic field reconnection during the SOL2014-09-10T17:45 flare}

\titlerunning {Magnetically coupled atmosphere}

\author{ H. M\'esz\'arosov\'a\inst{1}
    \and P. G{\"o}m{\"o}ry\inst{2}     }

\institute{ Astronomical Institute of the Academy of Sciences of the Czech Republic,
            CZ--25165 Ond\v{r}ejov, Czech Republic
       \and Astronomical Institute of the Slovak Academy of Sciences, SK--05960 Tatransk\'a Lomnica, Slovak Republic
       }

\offprints{H. M\'esz\'arosov\'a, \email{hana@asu.cas.cz}}

\date{Received ................ / Accepted ...................}

\abstract
{}
{We study the physical properties and behaviour of the solar atmosphere during the GOES~X1.6 solar flare on 2014~September~10.}
{The steady plasma flows and the fast sausage MHD waves were analysed with the wavelet separation method. The magnetically coupled atmosphere and the forced magnetic field reconnection were studied with the help of the Vertical-Current Approximation Non-linear Force-Free Field code.}
{We studied a~mechanism of MHD wave transfer from the photosphere without dissipation or reflection before reaching the corona and a~mechanism of the wave energy distribution over the solar corona. We report a~common behaviour of (extreme)ultraviolet steady plasma flows (speed of 15.3$\rightarrow$10.9~km~s$^{-1}$) and fast sausage MHD waves (Alfv\'en speed of 13.7$\rightarrow$10.3~km~s$^{-1}$ and characteristic periods of 1\,587$\rightarrow$1\,607~s), propagating in cylindrical plasma waveguides of the individual atmospheric layers (photosphere$\rightarrow$corona) observed by SDO/AIA/HMI and IRIS space instruments. A~magnetically coupled solar atmosphere by a~magnetic field flux tube above a~sunspot umbra and a~magnetic field reconnection forced by the waves were analysed. The solar seismology with trapped, leakage, and tunnelled modes of the waves, dissipating especially in the solar corona, is discussed with respect to its possible contribution to the outer atmosphere heating.}
{We demonstrate that a~dispersive nature of fast sausage MHD waves, which can easily generate the leaky and other modes propagating outside of their waveguide, and magnetic field flux tubes connecting the individual atmospheric layers can distribute the magnetic field energy across the active region. This mechanism can contribute to the coronal energy balance and to our knowledge on how the coronal heating is maintained.}

\keywords{Sun: flares -- Sun: atmosphere -- Sun: corona -- Sun: UV radiation -- Sun: oscillations -- Sun: magnetic fields}

\maketitle

\section{Introduction}
The coronae of the Sun and solar-type stars are considerably hotter than their photospheres. Even after the discovery of the million-degree coronal temperatures (Grotrian 1939; Edlen 1943), understanding the heating of the solar corona still remains unclear (maintaining a continual and a high amount of energy flux to compensate for coronal radiative losses). Moreover, the entire solar atmosphere is a~highly coupled system of different atmospheric layers and magnetic fields play a~key role in this coupling. Most of the contemporary theories for coronal heating postulate that free energy is injected from the photosphere into the corona where it is converted into heat through the reconnection of magnetic field lines or by wave dissipation. 

One of the possible mechanisms is coronal heating via a~large number of small impulsive heating events called nanoflares (e.g. Parker 1988; Ishikava et al. 2017; Asgari-Targhi et al. 2019). The impulsively heated high-temperature loops in active regions and a~study of their magnetic properties (Ugarte-Urra et al. 2019) have helped to reveal a~relationship between the magnetic field and coronal heating. Spicules, that is to say frequent small rapid plasma jets, observed in the solar chromosphere can occur at any moment and might supply hot plasma to the solar corona (Athay \& Holzer 1982; Samanta et al. 2019). They could provide a~link between magnetic activities in the lower atmosphere and coronal heating (De Pontieu et al. 2011; Klimchuk 2012). The maximum energy injected into the corona by spicules may largely balance the total coronal energy losses in quiet regions and possibly also in coronal holes, but not in active regions (Goodman 2014). Another possibility to transport energy between atmospheric layers is by Alfv\'en waves (Escande at al. 2019). Also, loop footpoint reconnection that is both inside and outside active regions (Reale et al. 2019; Wang at al. 2019) and a~forced magnetic reconnection triggered by external perturbation, for example by non-linear magnetohydrodynamic (MHD) waves (Hahm \& Kulsrud 1985), can contribute to the process of coronal heating. A~rapid forced magnetic field reconnection in the solar corona (Srivastava et al. 2019) presents a~novel physical scenario for the formation of temporary coronal reconnections, where plasma dynamics is forced externally by a~moving prominence.

Solar flare energy that accumulates in magnetic fields is explosively transformed into the plasma motions and heating, MHD waves, and the acceleration of particles. The solar atmosphere heating through wave dissipation (Arregui 2015) belongs to the traditional heating scenarios. The MHD waves (Aschwanden 2004) are among the important candidates for the mechanism of the energy transfer to the outer atmosphere and thus, they are critical for solving the coronal heating problem. The solar flare seismology (Stepanov et al. 2012; De Moortel \& Nakariakov 2012; Parnell \& De Moortel 2012; Soler et al. 2017) of a~magnetically coupled atmosphere with respect to the heating processes still shows problems, for example, the photospheric MHD waves can be an insufficient driver for coronal heating because of rapid dissipation and refraction before reaching the corona (Anderson \& Athay 1989). 

Among various MHD waves (Nakariakov \& Verwichte 2005), the highly dispersive impulsively generated, by various flare processes, fast sausage magnetoacoustic wave, which was theoretically predicted in Roberts at al. 1983, 1984, forms wave trains of short-, middle-, and long-period components that propagate along the waveguide (structure with enhanced plasma density). A~temporal evolution of these trains in a~dense slab waveguide, with a~plane geometry, exhibits specific wavelet power spectra with a~tadpole pattern (Nakariakov at al. 2004) consisting of a~narrow-spectrum tail preceding its broad-band head (that is, the tadpole goes tail-first). These wave trains that are studied in plasma cylindrical waveguides (Shestov et al. 2015) show this tadpole wavelet spectrum with either almost symmetrical head-tail shapes or shapes with the head preceding the tail (that is, the tadpole goes head-first). The abbreviation 'tadpole~wave' stands for the impulsively generated fast magnetoacoustic sausage wave train, propagating in a~plasma waveguide and displaying the characteristic tadpole pattern in the wavelet power spectrum of its time series. The term 'sausage' means axisymmetric initial pulse (the central axis remains undisturbed). The term 'fast' means that the wave train propagates significantly faster (M\'esz\'arosov\'a et al. 2014) than some characteristic speeds that can be estimated observationally, for example, the sound wave (Nakariakov \& Zimovets 2011). The slow magnetoacoustic waves do not create wave trains, that is to say no wavelet tadpole patterns are created. These slab tadpole waves were discovered in the 1999 solar-eclipse data (Katsiyannis et al. 2003) and they were detected, for example, in solar radio data (M\'esz\'arosov\'a et al. 2009a,b, 2016), extreme ultraviolet (EUV, SDO/AIA) emission intensity (Yuan et al. 2013), and in a~fan structure of the coronal magnetic null point (M\'esz\'arosov\'a et al. 2013). These studies were supported by MHD numerical simulations in, for example, Jel\'inek \& Karlick\'y (2012), Pascoe et al. (2013), and M\'esz\'arosov\'a et al. (2014). The cylindrical magnetic flux tubes with MHD waves were studied analytically (Edwin \& Roberts 1983; Moreels et al. 2013; Oliver at al. 2015) as well as the leakage modes of the MHD waves propagating outside the magnetic tube (Cally 1986; Yuan et al. 2011; Pascoe et al. 2013). 

This paper is organised as follows. The isolated sunspot observation is presented in Sect.~2. Steady plasma flows observed at individual atmospheric layers above the spot are analysed in Sect.~3. A~dispersive nature of fast sausage MHD waves accompanying these plasma flows are studied in Sect.~4. An evolution of the spot magnetic field strength and forced magnetic field reconnection are analysed in Sect.~5. Finally, a~discussion and conclusions are presented in Sect.~6. An appendix is dedicated to the description and explanation of the methods used.
\begin{figure}[ht]
\centering
\includegraphics[width=8.0cm]{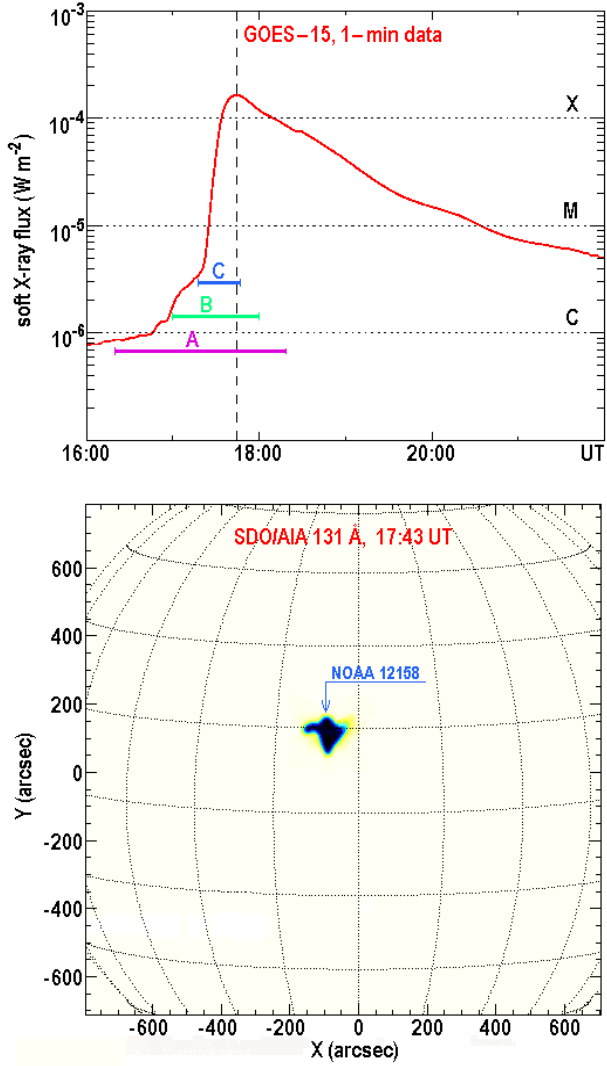}
\caption{$Upper~panel:$ GOES-15 soft X-ray flux reached its maximum at 17:45~UT (red curve).
The examined time interval 16:20--18:20~UT, the occurrence of the steady plasma flows (17:00--$\approx$18:00~UT), 
and the tadpole waves (17:19--17:48~UT) are indicated by bars A,~B, and~C, respectively.
$Bottom~panel:$ Location of the flare NOAA~12158 near the solar disc centre observed by SDO/AIA~131\AA~at 17:43~UT.}
\label{fig01}
\end{figure}
\begin{figure}[ht]
\centering
\includegraphics[width=9.0cm]{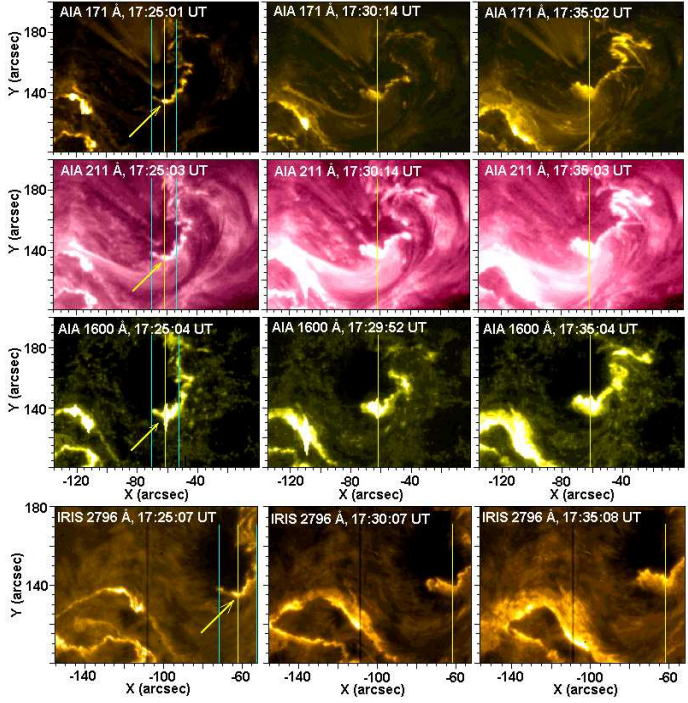}
\caption{Temporal examples of observed SDO/AIA 171, 211, 1600\AA, and IRIS 2796\AA~emissions above the sunspot. 
The artificial slit position of $X$~=~$-$62\arcsec, the IRIS spectrograph slit, and the positions of waveguide margins 
are displayed in yellow, black, and cyan vertical lines, respectively. 
Locations of tadpole wave triggering (bright blobs) are marked by arrows.}
\label{fig02}
\end{figure}
\begin{figure}[ht]
\centering
\includegraphics[width=9.0cm]{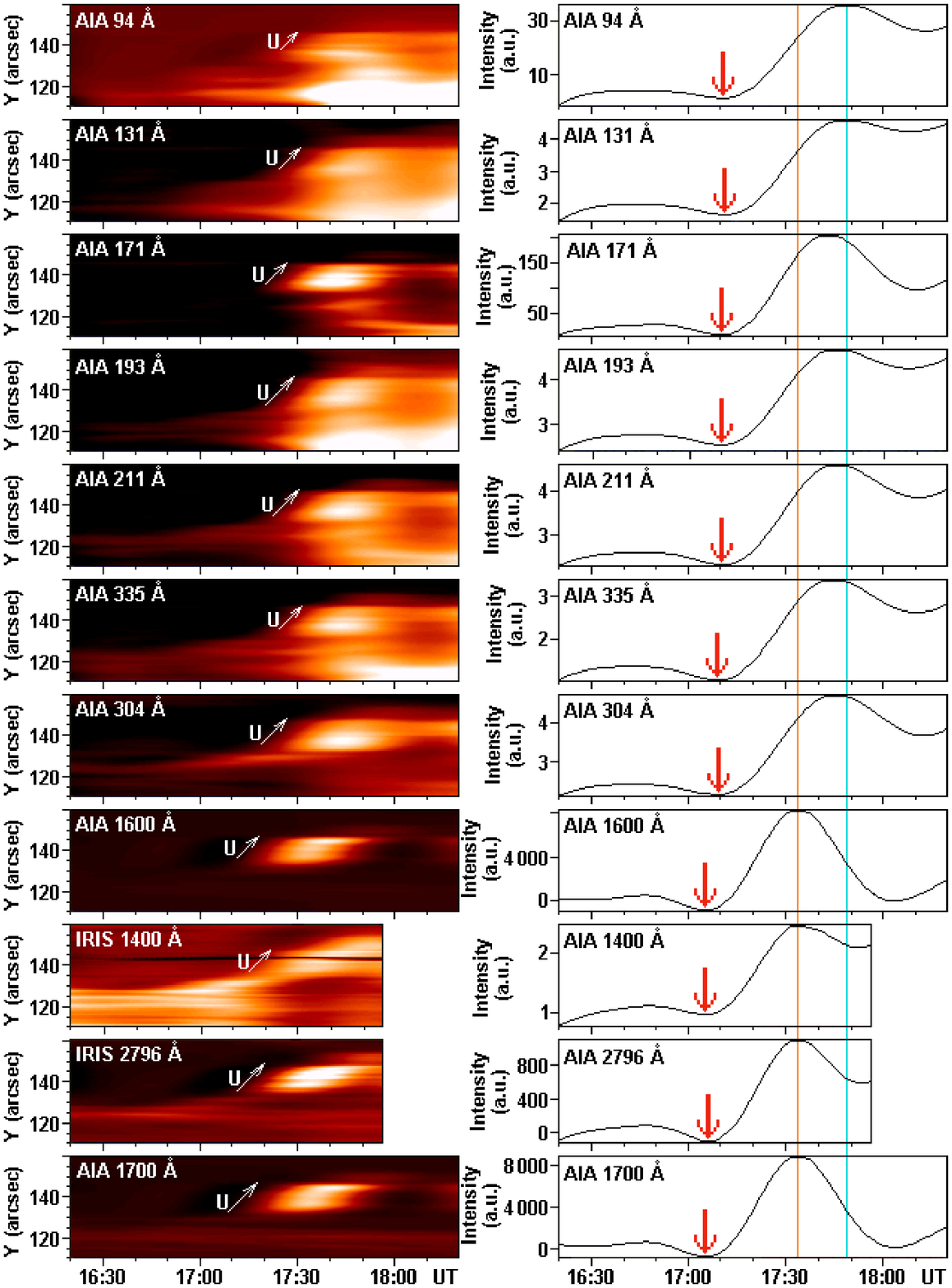}
\caption{Common behaviour of separated (E)UV steady plasma flows determined at the $X$~=~-62\arcsec~slit.
Flow speed~$U$ is marked by arrows. Vertical lines in selected flow time series at $Y$~=~139\arcsec (right panels) show the time of an emission maximum at AIA~1700 (orange) and 94\AA~observables (cyan). The red arrows show the starting times of the plasma flow onsets.}
\label{fig03}
\end{figure}
\section{Observation of the isolated sunspot SOL2014-09-10T17:45}
The isolated sunspot was observed on 2014 September 10 during the GOES-15 X1.6-class flare (SOL2014-09-10T17:45) by the following three instruments: (i)~the Atmospheric Imaging Assembly (AIA, Lemen et al. 2012) and (ii)~the Helioseismic and Magnetic Imager (HMI, Schou et al. 2012; Scherrer et al. 2012) on board the Solar Dynamics Observatory (SDO, Pesnell et al. 2012), and (iii) the Interface Region Imaging Spectrograph (IRIS, De Pontieu 2014). 

The SDO/AIA imaging data products of 94, 131, 171, 193, 211, 304, and 335\AA~channels (temporal cadence of 12~s) together with channels at 1600 and 1700\AA\  (cadence of 24~s)
were used with their spatial scale of 0.6\arcsec/pixel. The SDO/HMI data products consist of a~line-of-sight magnetogram (cadence of 45~s, spatial scale of 0.5\arcsec/pixel). The IRIS imaging data of the UV slit-jaw imaging 1400 and 2796\AA~channels with a~typical cadence of 18.8~s were used at the highest available spatial resolution of 0.166\arcsec/pixel. This flare being studied started at $\sim$16:47~UT and reached its GOES X-rays maximum at 17:45~UT (Fig.~1). The spot, which occurred in the active region NOAA 12158 and was located at the heliographic position N15W06 (near the solar disc centre, Fig.~1), was examined during the 16:20--18:20~UT interval (bar~A, Fig.~1). The IRIS observations finished at 17:58~UT and at the $X$~=~$-$50\arcsec coordinate. Selected samples of imaging data observed by the AIA~171, 211, 1600\AA, and IRIS 2796\AA~(Fig.~2) displayed their temporal evolution above the spot of positive polarity, which was surrounded by a~hooked ribbon. A~group of negative polarity pores surrounded by another ribbon were located at the spot SW. There was a~light bridge crossing the sunspot umbra centre at $Y\approx$146--152\arcsec.

\section{Analysis of the (E)UV steady plasma flows}
We studied the temporal and spatial evolution of all (E)UV data products. Since the strongest sunspot magnetic field ($ \approx$2\,200~G) was found along the $X$~=~$-$62\arcsec~coordinate, the space-time diagrams determined along this artificial slit (yellow lines, Fig.~2) were chosen for a~more detailed analysis. We used the WAvelet SEparation Method (WASEM) to separate the steady plasma flows contained in the original (E)UV space-time diagrams from other physical phenomena (Appendix~A.2). 

The space-time diagrams with these separated plasma flows in all the observables (Fig.~3) exhibited a~common behaviour. They displayed the flow temporal evolution with a~spatial drift~$U$ (arrows, left column). This drift was interpreted as the speed of the plasma flows decreasing with the solar altitude: $U$~=~15.38 and 10.99~km~s$^{-1}$ in the photosphere and the solar corona, respectively (Table~1).  
The plasma flows (Table~1) first appeared in the penumbral coordinates ($T_\mathrm{1}$=17:00--17:04~UT, $Y_\mathrm{1}$~=~132--134\arcsec) and later in the umbral ones ($T_\mathrm{2}$=17:11--17:19~UT, $Y_\mathrm{2}$~=~145--147\arcsec)
in all (E)UV observables.

The selected time series detected at $Y$~=~139\arcsec~(Fig.~3) show onsets with a~starting time of the plasma flows (red arrows) detected first in the inner atmosphere. The plasma flow intensity maximum ($T_\mathrm{peak}$, Table~1) of these onsets was observed during the time interval 17:33--17:48~UT, that is, at about the flare X-rays maximum (17:45 UT). The time difference between the 94\AA~(cyan vertical line) and the AIA 1700\AA~flow intensity maxima (orange vertical line) was found to be $\sim$16 minutes. 
The cross-correlation analysis of the individual (E)UV time series (reference 1700\AA~series) showed well correlated flows of individual atmospheric layers. An example of this analysis for the time series detected at $Y$~=~139\arcsec exhibited $CC_\mathrm{1700}$ coefficients of 0.72--0.99 (Table~1). The correlation maximal time shift $TS_\mathrm{1700}$ was found between the solar photosphere and corona (876~s = $\sim$15 minutes). Thus, these plasma flows were generated sooner in the inner atmosphere than in solar corona. The plasma flow distance $S=Y_\mathrm{2}-Y_\mathrm{1}$ was determined to be the largest in the photosphere (10\,150~km) and the plasma flow duration, $N_\mathrm{1}=T_\mathrm{2}-T_\mathrm{1}$, increased with the solar altitude (660 and 870~s for the photosphere and solar corona, respectively, Table~1).

These types of space-time diagrams ($Y$~=~110--160\arcsec, 16:30--18:00~UT) were analysed along individual artificial slits in the range 
of $X$~=~0~--~$-$140\arcsec~(SDO) and 
$X$~=~$-$50~--~$-$140\arcsec~(IRIS). Among them, the space-time diagrams displaying the common behaviour of the (E)UV observables were detected at all artificial slits in the range of $X$~=~$-$50~--~$-$70\arcsec. The plasma flow speed~$U$ value was the highest at space-time diagrams at the $X$~=~$-$62\arcsec~slit. This speed gradually decreased along with an increasing distance from this position.

\begin{table*}[h]
\caption{Characteristic parameters of (E)UV plasma flows.}
\label{tab1}
\centering
\begin{tabular}{ccccccccccc}
\hline\hline
Channel &\multicolumn{2}{c}{Plasma}       &\multicolumn{2}{c}{Plasma}       & Distance & Duration       & Plasma        & Time                  & $^b$Coefficient    & Time               \\
        &\multicolumn{2}{c}{flow start}   &\multicolumn{2}{c}{flow end}     &          &                & flow speed    &  {Y = 139\arcsec}     &                    & shift              \\
        & T$_\mathrm{1}$ & Y$_\mathrm{1}$ & T$_\mathrm{2}$ & Y$_\mathrm{2}$ &  S       & N$_\mathrm{1}$ &   U           & $^a$T$_\mathrm{peak}$ & CC$_\mathrm{1700}$ & TS$_\mathrm{1700}$ \\
$[$\AA] &    [UT]        & [\arcsec]      &     [UT]       &  [\arcsec]     & [km]     &  [s]           & [km~s$^{-1}$] & [UT]                  &                    & [s]                \\

\hline
        &                     \multicolumn{10}{c}{Corona}                                                                                                                                                                                                                                                                      \\
 94     & 17:03   & 132     & 17:19 & 146   & 10\,150  &  960  &  10.57  & 17:48   &   0.77  &   876       \\
 131    & 17:03   & 132     & 17:19 & 146   & 10\,150  &  960  &  10.57  & 17:48   &   0.72  &   872       \\
 171    & 17:04   & 133     & 17:17 & 145   &  8\,700  &  780  &  11.15  & 17:43   &   0.88  &   516       \\
 193    & 17:03   & 133     & 17:17 & 146   &  9\,425  &  840  &  11.22  & 17:45   &   0.73  &   756       \\
 211    & 17:04   & 134     & 17:17 & 146   &  8\,700  &  780  &  11.15  & 17:45   &   0.79  &   708       \\
 335    & 17:02   & 133     & 17:17 & 147   & 10\,150  &  900  &  11.28  & 17:44   &   0.79  &   636       \\
 \hline
 mean   &         &         &       &       &  9\,546  &  870  &  10.99  &         &         &             \\
 \hline
        &                    \multicolumn{10}{c}{Chromosphere and Transition Region}                                                                            \\
 304    & 17:01   & 133     & 17:13 & 146   &  9\,425  &  720  &  13.09  & 17:44   &   0.82  &   648       \\
 1400   & 17:02   & 134     & 17:15 & 146   &  8\,700  &  780  &  11.15  & 17:36   &$^c$0.91 &   266       \\
 1600   & 17:01   & 133     & 17:12 & 147   & 10\,150  &  660  &  15.38  & 17:33   &   0.99  &   0         \\
 2796   & 17:02   & 133     & 17:15 & 146   &  9\,425  &  780  &  12.08  & 17:35   &$^c$0.96 &   190       \\
 \hline
 mean   &         &         &       &       & 9\,425   &  735  &  12.92  &         &         &             \\
 \hline
        &                   \multicolumn{10}{c}{Photosphere}                                                                                                             \\
 1700   & 17:00   & 133     & 17:11 & 147   & 10\,150  &  660  &  15.38  & 17:33   &         &              \\
\hline
\end{tabular}
\tablefoot{$^a$T$_\mathrm{peak}$ = time of plasma intensity maximum detected in time series measured at $Y$ = 139\arcsec,  
           $^b$Coefficient = cross-correlation coefficient. (Note: We doubled each value of the AIA 1600 and 1700\AA~time 
           series to get them with a 12~s cadence instead of the original 24~s. $^c$The 1700 time series length and cadence  
           were adapted to the IRIS time series.)}
\end{table*}

\begin{table*}[h]
\caption{Characteristic parameters of (E)UV tadpole waves - part I.}
\label{tab2}
\centering
\begin{tabular}{cccccccccccccc}
\hline\hline
Channel & \multicolumn{2}{c}{Start} & \multicolumn{2}{c}{End} & Cylindrical    & Duration & Alfv\'en       & $^a$Radius     & Phase & \multicolumn{3}{c}{Period}                                         \\
        &       &           &                &                &   length       &          &  speed         &                & speed               &                 &                  &                 \\
        & T$_\mathrm{3}$    & Y$_\mathrm{3}$ & T$_\mathrm{4}$ & Y$_\mathrm{4}$ &   L      & N$_\mathrm{2}$ & v$_\mathrm{A}$ & R & v$_\mathrm{ph}$ & P$_\mathrm{ch}$ & P$_\mathrm{max}$ &P$_\mathrm{min}$ \\
$[$\AA] & [UT]  & [\arcsec] & [UT]           & [\arcsec]      &  [km]          &  [s]     & [km s$^{-1}$]  &  [km]          & [km s$^{-1}$]       & [s]             & [s]              & [s]             \\
\hline
        &                            \multicolumn{12}{c}{Corona}                                                                                                                                     \\
  94    & 17:25 & 132 & 17:43 & 147  & 10\,875  &  1\,080  &  10.07   & 5\,688  &  14.70  & 1\,480   & 1\,899  & 209  \\
 131    & 17:22 & 132 & 17:39 & 147  & 10\,875  &  1\,020  &  10.66   & 6\,457  &  13.70  & 1\,587   & 2\,130  & 214  \\
 171    & 17:19 & 130 & 17:37 & 145  & 10\,875  &  1\,080  &  10.07   & 6\,100  &  13.70  & 1\,587   & 2\,130  & 197  \\
 193    & 17:21 & 131 & 17:37 & 145  & 10\,150  &   960    &  10.57   & 6\,862  &  11.93  & 1\,701   & 2\,256  & 160  \\
 211    & 17:20 & 132 & 17:38 & 147  & 10\,875  &  1\,080  &  10.07   & 6\,100  &  13.70  & 1\,587   & 2\,106  & 158  \\
 335    & 17:21 & 132 & 17:38 & 147  & 10\,875  &  1\,020  &  10.66   & 6\,921  &  12.79  & 1\,701   & 2\,155  & 180  \\
 \hline
 mean   &       &     &       &      & 10\,754  &  1\,040  &  10.35   & 6\,355  &  13.42  & 1\,607   & 2\,113  & 186  \\
 \hline
        &                         \multicolumn{12}{c}{Chromosphere and Transition Region}                                                                                                       \\
 304    & 17:21 & 132 & 17:37 & 147  & 10\,875  &   960    &  11.33   & 7\,356  &  12.79  & 1\,701   & 2\,105  & 170  \\
 1400   & 17:20 & 131 & 17:37 & 146  & 10\,875  &  1\,020  &  10.66   & 6\,152  &  14.38  & 1\,512   & 2\,050  & 152  \\
 1600   & 17:25 & 132 & 17:48 & 156  & 17\,400  &  1\,380  &  12.61   & 7\,638  &  21.93  & 1\,587   & 2\,138  & 474  \\
 2796   & 17:22 & 131 & 17:45 & 153  & 15\,950  &  1\,380  &  11.56   & 6\,226  &  22.61  & 1\,411   & 1\,614  & 160  \\
 \hline
 mean   &       &     &       &      & 13\,775  &  1\,185  &  11.54   & 6\,843  &  17.93  & 1\,553   & 1\,977  & 239  \\
 \hline
        &                           \multicolumn{12}{c}{Photosphere}                                                                                                                               \\
 1700   & 17:24 & 131 & 17:46 & 156  & 18\,125  &  1\,320  &  13.73   & 8\,317  &  22.84  & 1\,587   & 2\,138  & 317  \\
\hline
\end{tabular}
\tablefoot{$^a$Radius = cylindrical cross-section radius}
\end{table*}

\begin{table}[h]
\caption{Characteristic parameters of (E)UV tadpole waves - part II.}
\label{tab3}
\centering
\begin{tabular}{ccccc}
\hline\hline
Channel & Mach   & $^a$NPS & $^b$Coefficient & Time                                     \\
        & number &         &                 & shift                                    \\
        & M$_\mathrm{0}$   & a$_\mathrm{ph}$ & CC$_\mathrm{1700}$ & TS$_\mathrm{1700}$  \\
$[$\AA] &        &         &                 & [s]                                      \\
\hline
        &          \multicolumn{4}{c}{Corona}                          \\ 94     & 1.05   & 1.32    &   0.82         &  192                     \\ 131    & 0.99   & 1.18    &   0.85         &  156                     \\ 171    & 1.11   & 1.22    &   0.89         &  180                     \\ 193    & 1.06   & 1.03    &   0.90         &  108                     \\ 211    & 1.11   & 1.22    &   0.89         &  108                     \\ 335    & 1.06   & 1.09    &   0.90         &   60                     \\ \hline
 mean   & 1.06   & 1.18    &                &                          \\ \hline
        &      \multicolumn{4}{c}{Chromosphere and Transition Region}  \\ 304    & 1.15   & 1.02    &   0.87         &   48                     \\ 1400   & 1.05   & 1.23    &   $^c$0.87     &   95                     \\ 1600   & 1.22   & 1.59    &   0.98         &    0                     \\ 2796   & 1.04   & 1.79    &   $^c$0.80     &  152                     \\ \hline
 mean   & 1.12   & 1.42    &                &                          \\ \hline
        &         \multicolumn{4}{c}{Photosphere}                      \\ 1700   & 1.12   & 1.54    &                &                          \\\hline
\end{tabular}
\tablefoot{$^a$NPS = normalised phase speed \\
           $^b$Coefficient = cross-correlation coefficient. (Note: We doubled each value of the AIA 1600 and 1700\AA~time series 
           to get them with 12~s cadence instead of the original 24~s. $^c$The 1700 time series length and cadence  
           were adapted to the IRIS time series.)}
\end{table}

\begin{figure}[h]
\centering
\includegraphics[width=9cm]{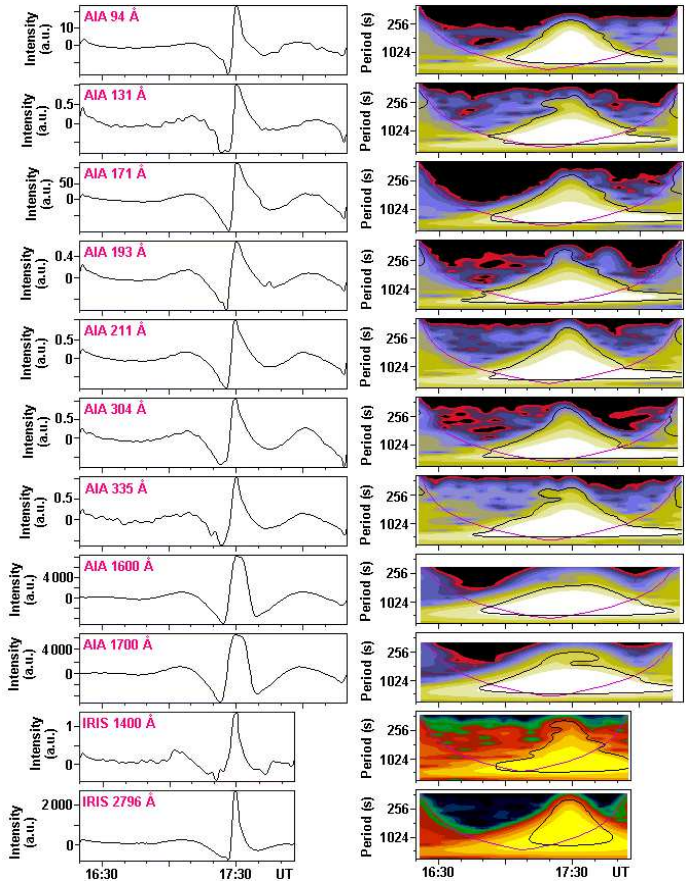}
\caption{Common behaviour of cylindrical tadpole waves in (E)UV observables determined at the $X$~=~-62\arcsec~slit. Selected time series observed 
at $Y$~=~139\arcsec (left column) and their corresponding wavelet power spectra (right column) with the cylindrical tadpole wave patterns (lighter areas) of 99\% significance (black contours).}
\label{fig04}
\end{figure}
\begin{figure}[ht]
\centering
\includegraphics[width=7cm]{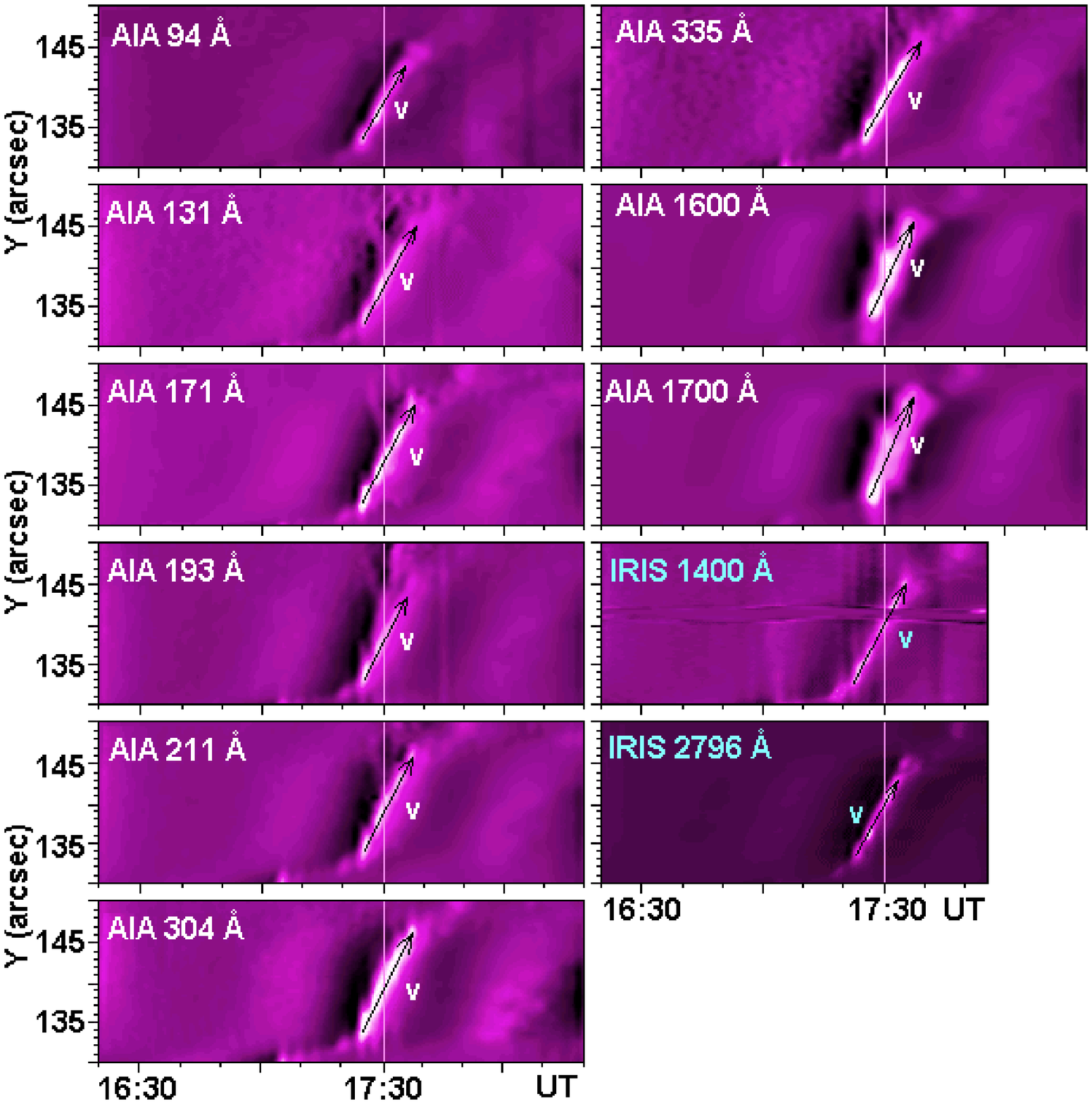}
\caption{Dynamical spectra of the cylindrical tadpole waves determined at the $X$~=~$-$62\arcsec~slit. The waves propagated with the highest speed~$v$ (arrows).}
\label{fig05}
\end{figure}
\section{Analysis of the (E)UV cylindrical tadpole waves}
All time series on individual $Y$~coordinates of the original (E)UV space-time diagrams of the $X$~=~$-$62\arcsec~slit were studied by the WASEM analysis (Appendix~A, Fig.~A.2) to search for signatures of tadpole waves in their wavelet power spectra. These tadpole waves were found as they show a~common behaviour for all (E)UV observables. Selected wave power patterns determined on the $Y$~=~139\arcsec~coordinate (Fig.~4) displayed either almost symmetrical head-tail shapes (e.g. IRIS 2796\AA) or shapes with a~head preceding the tail (e.g. AIA 171\AA). These types of wavelet patterns belong to tadpole waves propagating in a~cylindrical waveguide (Shestov et al 2015). They were detected during an interval of $\approx$17:19--17:46~UT ($T_\mathrm{3}$, $T_\mathrm{4}$, Table~2). The coordinates $Y_\mathrm{3}$ and $Y_\mathrm{4}$ displayed a~spatial interval of 130--156\arcsec, that is to say these waves propagated in a similar way as the plasma flows, in other words, from the penumbra coordinates towards the umbra light bridge ones. Both the waveguide cylinder length $L = Y_\mathrm{4} - Y_\mathrm{3}$ and the wave duration $N_\mathrm{2} = T_\mathrm{4} - T_\mathrm{3}$ decrease with the solar altitude, for example, $L = $18.1 \& 10.7~Mm and $N = $1\,320 \& 1\,040~s in the photosphere and the solar corona, respectively.

Dynamic spectra of all tadpole wave time series were computed for the $Y$~=~130--150\arcsec~coordinates and the highest wave speed~$v$ was determined (arrows, Fig.~5). It can be used as a~waveguide internal Alfv\'{e}n speed~$v_\mathrm{A}$ approximation (Nakariakov et al. 2004). The Alfv\'en speed $v_\mathrm{A} = B_\mathrm{0}/\sqrt{\mu_\mathrm{0} \rho_\mathrm{0}}$, where $B_\mathrm{0}$ is the waveguide internal magnetic field strength, $\mu_\mathrm{0}$ is the vacuum permeability, and $\rho_\mathrm{0}$ is the plasma density (Nakariakov et al. 2012). Thus, the speed $v_\mathrm{A}$ might be proportional to the magnetic field $B_\mathrm{0}$, which is strongest in the photosphere and weaker in the corona). Therefore, the highest speed $v_\mathrm{A}$ (Table~2) was found in the photosphere (13.73~km~s$^{-1}$) and the lowest one was found in the corona (10.35~km~s$^{-1}$).

The main feature of the dispersive mechanism for a~tadpole wave formation is a~period modulation (Roberts at al. 1984). We determined characteristic ($P_\mathrm{ch} \sim$ 1\,600~s), maximal ($P_\mathrm{max} \sim$ 2\,000~s), and minimal ($P_\mathrm{min}$) wave periods with WASEM (Appendix~A.2). The $P_\mathrm{min}$ periods changed at individual $Y$-coordinates according to the tadpole head fins (Shestov et al 2015) evolution. This $P_\mathrm{min}$ period decreased from the photosphere ($P_\mathrm{min}$=317~s) towards the solar corona (mean $P_\mathrm{min}$=186~s). Thus, the highest period modulation ($P_\mathrm{max}$--$P_\mathrm{min}$) was found in the corona (Table~2). 

A~waveguide cross-section radius of $R\approx P_\mathrm{ch} \cdot v_\mathrm{A}/2.62$ (Nakariakov et al. 2003) was computed for all (E)UV observables (Table~2). For example, we determined the period $P_\mathrm{ch}$ = 1,587~s and the Alfv\'en speed $v_\mathrm{A}$ = 10.07~km~s$^{-1}$ for the AIA 211\AA~(Table~2). Then the cylinder radius~$R$ was equal to 6,100~km, that is, $\sim$~8\arcsec. The cylinder margins ($X=-70\arcsec$ and $-54\arcsec$, i.e. $-$62$\mp$8\arcsec) displayed the waveguide width of~2$R$ (vertical cyan lines, first column, Fig.~2) . The wave phase speed $v_\mathrm{ph} = \omega/k = 2L/P_\mathrm{ch}$ (where $\omega = 2\pi/P_\mathrm{ch}$, longitudinal wavenumber $k = \pi/L$) satisfied the waveguide condition (Pascoe at al. 2007) $v_\mathrm{ph}>v_\mathrm{A}$ for all observables.

The cross-correlation analysis (reference 1700\AA~time series) showed well correlated individual (E)UV time series in the individual atmospheric layers. 
A~characteristic example of this analysis for the time series detected at $Y$~=~139\arcsec~exhibited $CC_\mathrm{1700}$ coefficients of 0.82--0.98 (Table~3). The correlation maximal time shift $TS_\mathrm{1700}$ was found between the photosphere and corona (192~s). Thus, the tadpole waves appeared in the inner atmosphere sooner, by $\sim$3 minutes, than in the solar corona. 

A~dimensionless phase speed~$a_\mathrm{ph}$ (Table~3) of individual tadpole waves were normalised on the waveguide internal Alv\'en speed (Nakariakov \& Roberts 1995): $a_\mathrm{ph} = \omega/k (v_\mathrm{A} + M_\mathrm{0})$, where $M_\mathrm{0} = U/v_\mathrm{A}$ is the Mach number inside the cylinder and $U$~is the plasma flow speed (Table~1). The speed $a_\mathrm{ph}$ of individual tadpole waves ranged from 1.02--1.79 and was the highest in the inner atmosphere and decreased towards the solar corona. 

Generally, tadpole waves are impulsively triggered by a~perturbation in the radial velocity, localised at the cylinder axis (Shestov et al. 2015). Therefore, we searched for a~possible initial impulse in all (E)UV data products. We found it on coordinates of $X$~=~$-$62\arcsec, $Y\sim$132\arcsec~at 17:25~UT. It was exhibited as a~bright blob of spatial size $\Delta X \sim \Delta Y \sim$2~Mm (arrows, Fig.~2). 

\begin{figure*}[h]
\centering
\includegraphics[width=17cm, height=16cm]{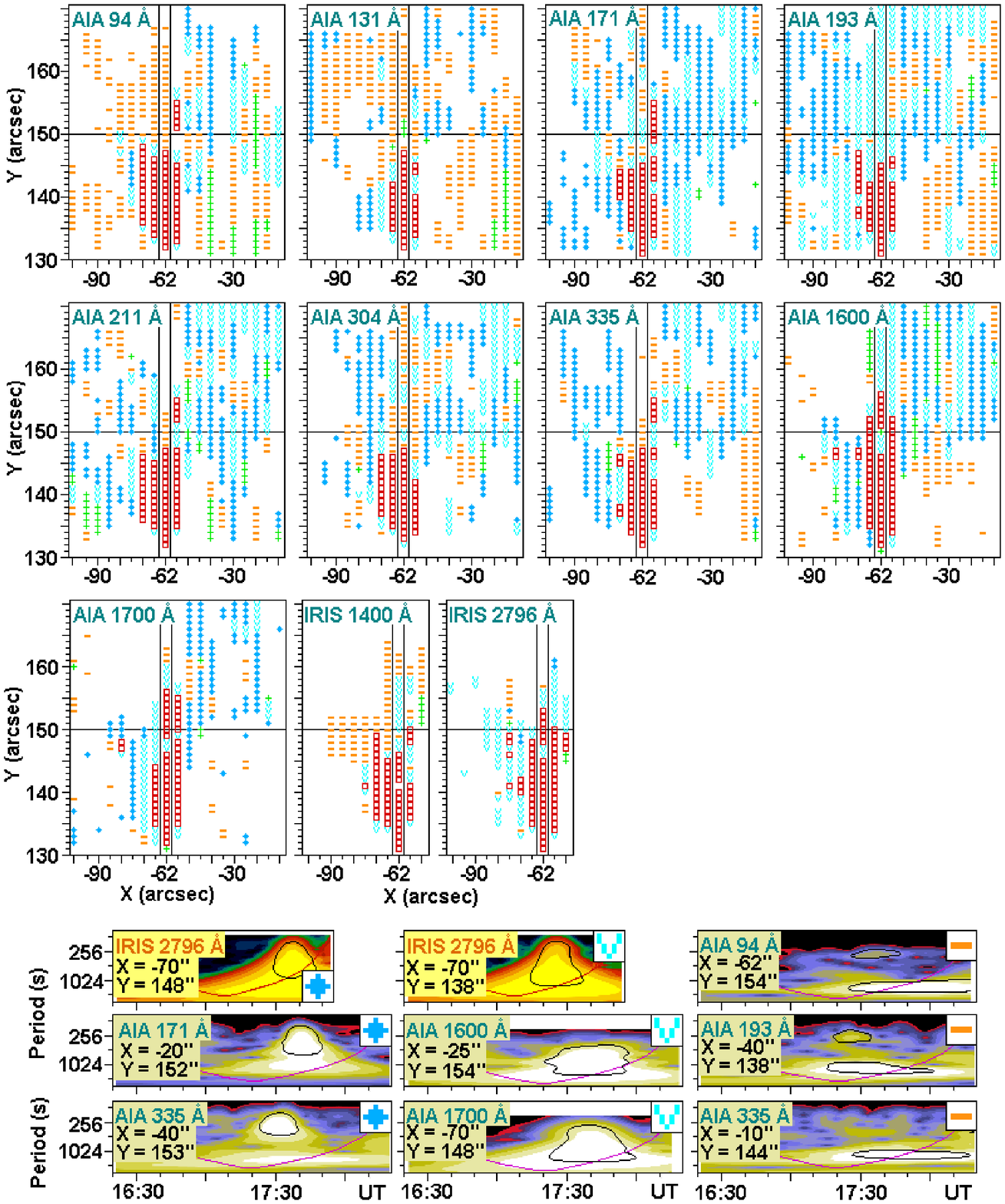}   
\caption{Distribution diagrams of trapped (red squares), leakage (orange bars), tunnelled (blue stars), reduced (cyan arches), and plane slab (green crosses) tadpole waves of individual (E)UV observables (upper part) with characteristic wavelet power examples (bottom part). Vertical lines show the slit position of $X$~=~$-$62\arcsec~(instead of $X$~=~$-$60\arcsec).}
\label{fig06}
\end{figure*}
\subsection{Trapped and leakage tadpole wave modes}
The tadpole wave is highly dispersive with a~sequence of expansions and contractions of the cylindrical cross-section radius accompanied by a~variation in the plasma density and the absolute value of the magnetic field (Nakariakov et al. 2004; Stepanov et al. 2012). We studied individual tadpole wave modes in a~selected part of the active region related to the sunspot with WASEM. With respect to the two-hour observation interval, we determined these modes in the $Y$~range of 130--170\arcsec~(with 1\arcsec~step) and for the slit range of $X = -$10~--~$-$100\arcsec~(with a 5\arcsec~step). The $X$~=~$-$62\arcsec~slit (cylinder axis) is presented instead of the $X$~=~$-$60\arcsec~slit. Empty places in these diagrams mean there is no significant wavelet pattern  nor a~complex wavelet signature with an unclear interpretation. The modes of tadpole waves were divided into five distinct groups: trapped, leakage, tunnelled, reduced, and slab tadpole waves. They are displayed in distribution wave mode diagrams (Fig.~6) for the individual (E)UV observables with some characteristic examples in their wavelet power spectra:  
(i) firstly, the trapped tadpole wave (red squares) remained localised close to the cylinder axis of $X$~=~$-$62\arcsec (wavelet examples, Fig.~4). These trapped waves experience an internal reflection at the cylinder surface and are evanescent outside their waveguides (Nakariakov at al. 2012). 
(ii) Secondly, the leakage tadpole wave mode (orange bars) leaked from a~mode waveguide to propagate out of the cylinder (Nakariakov at al. 2012) at the external Alfv\'en speed $v_\mathrm{Ae}$, which can be approximated by the phase speed $v_\mathrm{ph}$ (Table~2). The leakage periods are always longer than the trapped ones. The tadpole wave becomes leaky for wavelengths greater than a~cutoff, that is, the ratio of the wavelength $\lambda = 2\pi/k$ and the speed $v_\mathrm{Ae}$ (Pascoe et al. 2007). This cutoff increases with steepness and the plasma density (or Alfv\'{e}n speed) contrast ratio. The highest approximated Alfv\'en speed contrast ratio was found in the inner atmosphere (Table~2). We assumed the cutoff period $P_\mathrm{c} \ga P_\mathrm{ch}$ and observed that leakage periods~$P$ ranged from $P_\mathrm{c} \la P < P_\mathrm{max}$. These leaky waves propagated over a~distance of $\Delta~Y \approx$1--2\arcsec, that is, they decayed rather quickly. 
(iii) Thirdly, the tadpole waves in a~tunnelled regime (blue stars) may occur in another (parallel) waveguide close the original one (Ogrodowczyk \& Murawski 2007). These tunnelled waves propagate in a~trapped regime and depend on the parameters for both the original and the new waveguide. The short-period wave components dominate with an increasing distance between these waveguides. Generally, the narrower waveguide enables the propagation of tadpole waves with just shorter periods (M\'{e}sz\'{a}rosov\'{a} et al. 2014). Only short-period components of the tunnelled waves were found, and their periods~$P$ ranged from $P_\mathrm{min} \la P \ll P_\mathrm{ch}$ (Table~2). The individual tunnelled waves propagated over a~distance of $\Delta~Y \approx$5--10\arcsec~, supporting the idea of trapped wave modes in their new waveguide (they are not so quickly dumped as the leakage modes). 
(iv) Fourthly, we found reduced tadpole wave modes where mainly medium-period components of the original trapped waves were propagated (cyan arches, Fig.~6). They were detected in locations with less suitable conditions for the trapped wave propagation
(e.g. the beginning, ends, and margins of the waveguide).  
(v) Fifthly, the slab tadpole waves (green crosses) mean a~planar (two-dimensional) geometry where the tadpole tail precedes its head. Whereas the real cylindrical waveguides (three-dimensional) can have a~rather circular cross-section profile with axes $R1 \approx R2$, and the planar slab tadpole wave (Katsiyannis et al. 2003; M\'esz\'arosov\'a et al. 2009a,b, 2014, 2016) can be realised in a~heavily flattened cylinder (with, e.g. $R1 << R2$). 

These leakage, tunnelled, reduced, and slab tadpole waves can propagate in different directions outside the original waveguide according to local physical conditions in their new locations. The most frequent occurrence of these waves was detected in the corona (Fig.~6), where a~large number of convenient new waveguides (e.g. coronal loops) were close to the sunspot.
\begin{figure*}[h]
\centering
\includegraphics[width=18cm]{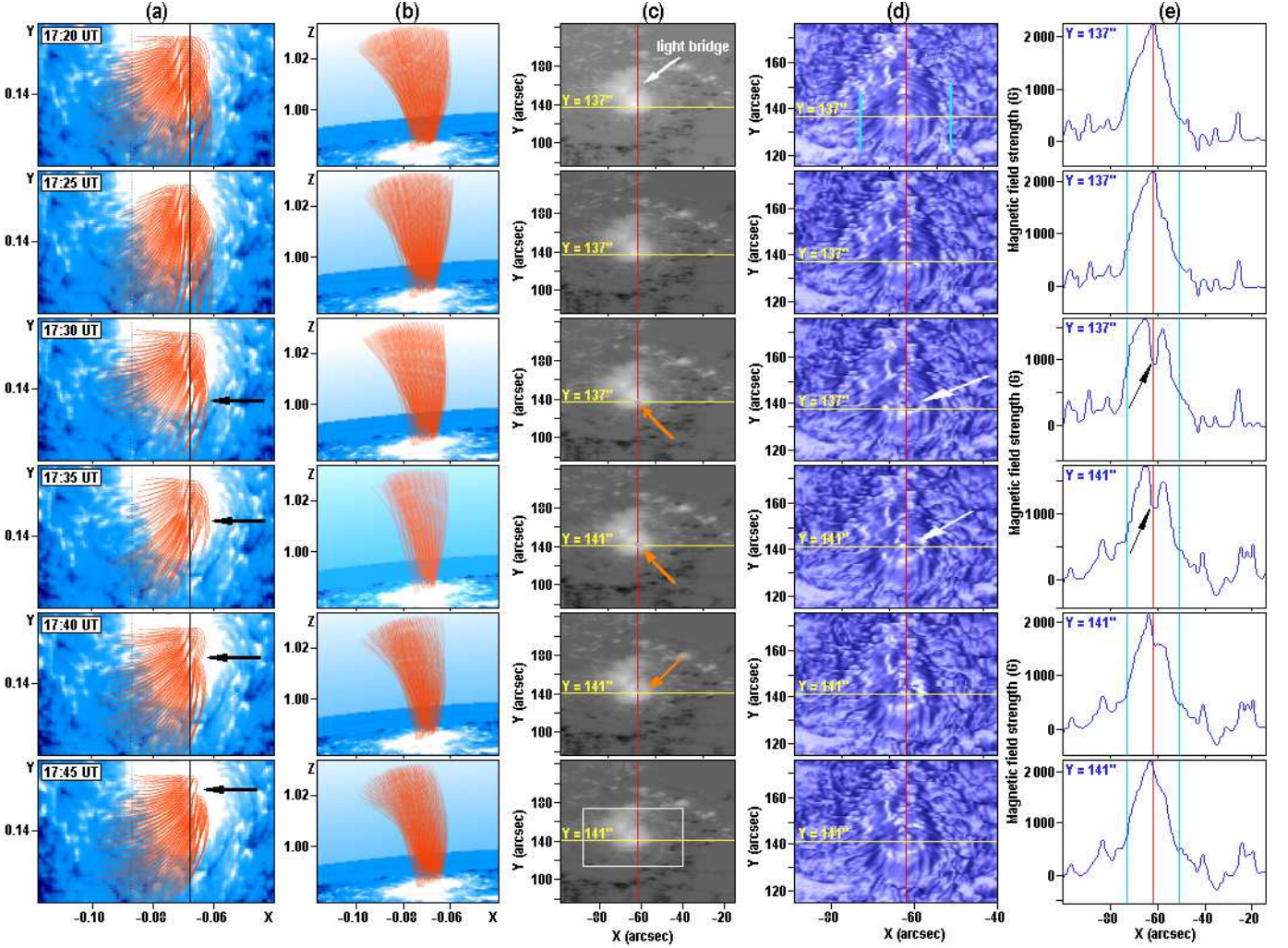}
\caption{Magnetically coupled atmosphere and magnetic field reconnection forced by passing tadpole wave (UT times $(a)$ are related to individual rows of plots). Magnetic field perturbations are marked by arrows (except 17:20 UT, $(c)$). The slit position $X$~=~$-$62\arcsec~is displayed by vertical red lines. 
$(a)$--$(b)$: The blue background represents the $B_\mathrm{z}$ HMI magnetograms. The best-fit magnetic field lines (orange curves) were computed for AIA 1600\AA~with the VCA-NLFFF code: Footpoints are in a~150~$\times$~150 grid, field strength of $|B_\mathrm{z}|>$180~G. The field lines were obtained from line-of-sight angles ($(a)$) and the 90~deg rotated to north angles ($(b)$). 
$(c)$--$(e)$: Evolution of magnetic field perturbation in SDO/HMI magnetograms. Dark mottles (orange arrows, $(c)$) show that places of magnetic field strength decrease.  
The magnetic field strength distributions ($(d)$) exhibit a~hill shape. They are related to the area size (white rectangle, $(c)$).
The magnetic field strength profiles ($(e)$) are at selected positions (yellow horizontal lines, $(c)$--$(d)$).
The SDO/AIA 1700\AA~waveguide width is marked by vertical cyan lines ($(d)$--$(e)$).}
\label{fig07}
\end{figure*}
\section{Magnetically coupled atmosphere and forced magnetic field reconnection}
We used (i) the Vertical-Current Approximation Non-linear Force-Free Field (VCA-NLFFF) code (Aschwanden 2016) to exhibit a~temporal evolution of the vertical non-linear force-free magnetic field in the solar atmosphere above the sunspot (the VCA-NLFFF parameters that were used are explained in Appendix~B). We also used (ii) the SDO/HMI magnetograms to explore the magnetic field strength evolution above the sunspot. 

The best-fit magnetic field lines (orange curves, $(a)$--$(b)$, Fig.~7)) were computed with footpoints in a~150~$\times$~150 rectangular grid with the VCA-NLFFF. The highest magnetic field strength was selected ($|B_\mathrm{z}|>$180~G) to exhibit only magnetic field lines around the $X$~=~$-$62\arcsec~slit (vertical red lines, $(a)$).
The blue background image represents the $B_\mathrm{z}$~HMI magnetogram with the sunspot (in white). The magnetic field lines were computed from two different aspect angles (Aschwanden 2016): the line-of-sight $X$--$Y$ plane of the solar surface ($(a)$) and the 90~deg rotated to the north ($(b)$, $Z$~=~altitude). The coordinates~$X$, $Y$, and~$Z$ are expressed in solar radii. These magnetic field lines exhibited a~connection between the inner atmosphere and the solar corona. This magnetic flux tube was localised above the sunspot and caused a~magnetically coupled solar atmosphere by the strong magnetic field ($\approx$2\,200~G). All SDO/AIA observables were computed separately with similar results. Therefore, AIA 1600\AA~magnetic field lines are presented as a~characteristic example ($(a)$--$(b)$). The individual time moments (17:20, 17:25, 17:30, 17:35, 17:40, and 17:45~UT) are related to the individual rows of plots. 

An evolution of the magnetic field perturbation in the SDO/HMI magnetograms is exhibited ($(c)-(e)$, Fig.~7) with the $X$~=~$-$62\arcsec~slit position (vertical red lines).
The light bridge location is shown (arrow, 17:20~UT). Detail distributions of the magnetic field strength ($(d)$) above the spot umbra are related to the area size displayed by the white rectangle ($(c)$, 17:45~UT). This area size is the same for all $(d)$~panels. These distributions show a~hill shape with steep slopes and a remarkable magnetic field maximum ($B_\mathrm{max}$) around the slit. The magnetic field strength coordinate (from the range 0--3\,000~G uniformly) is perpendicular to the figure plane (not displayed). Individual magnetic fluxes (yellow horizontal lines, $(c)$--$(d)$) were selected to display the evolution of the magnetic field strength ($(e)$). We determined the cylinder waveguide cross-section radius~$R$ in the photosphere (AIA 1700\AA, Table~2). The cylinder margin locations (vertical cyan lines, columns $(d)$--$(e)$) matched the area of the highest magnetic field strength well ($(e)$).

The 17:20~UT plots (Fig.~7) show the~situation before the tadpole wave occurrence, that is, before the perturbation of magnetic field lines. The VCA-NLFFF plots exhibited many magnetic field lines, which were open or declined towards the negative polarity pores ($(a)$). They created the characteristic dense magnetic field canopy ($(b)$). The HMI plots displayed an undisturbed magnetic field with a hill shape ($(d)$) of the $\approx$2\,200~G magnetic field strength ($(e)$) around the slit position. This magnetic field was stable without any significant changes during 16:59--17:23~UT. 
In the 17:25~UT plots, the tadpole wave occurred at the penumbra and the umbra magnetic field was only slightly disturbed. The 17:30--17:40~UT plots exhibit the consequences of the tadpole wave propagation over the umbra towards the light bridge. The original magnetic field lines ($(a)-(b)$) have become sparse and many are missing. It means that the passing tadpole wave caused a~reconnection among the individual magnetic field lines in places marked by arrows ($(a)$). The positions of propagating magnetic field perturbations were visible as small dark mottles (orange arrows, $(c)$), exhibiting a~decreased magnetic field strength in the spot. The perturbations caused local drops in the hill (arrows, $(d)$). This decreased magnetic field was well visible at selected $Y$~=~137/141\arcsec~fluxes ($(e)$) as a~decrease in the original magnetic field strength 
($\approx$2\,200~G) up to the new maximal value of $\approx$1\,700~G). The local minimum of $\approx$1\,000~G occurred just at the slit position (arrows, $(e)$). When the tadpole wave left the individual place above the umbra, the magnetic field lines were gradually recovered there (17:45~UT ($(b)$). The maximal magnetic field strength reached its original value of $\approx$2\,200~G around the slit once again (17:45, $(e)$).

Furthermore, we studied a~temporal evolution of extrapolated coronal emission measure~$EM$ and temperature~$T$ (Aschwanden et al. 2013) of the sunspot region. The HMI magnetograms and coronal AIA observations (94, 131, 171, 193, 211, and 335\AA) were used to construct the $EM$ and~$T$ spatial distributions. Characteristic AIA filtergrams at 17:01, 17:24, and 17:25~UT were selected (Fig.~8). The peak emission measure~$EM$ maps represent the total mass distribution of plasma given by the sum of the squared electron density of plasma. The temperature~($T$) maps only represent the most dominant temperature, characterising the plasma with the highest emission measure ($EM$) along the line of sight. 

The spot ribbon that is weakly visible at 17:01~UT was activated in 17:24~UT. A~bright blob was detected in 17:25~UT at the penumbral (arrows, $Y$~=~132\arcsec) slit position (vertical lines, $X$~=~$-$62\arcsec). The same bright blob was observed in the original spatial maps (arrows, Fig.~2, 17:25~UT). The characteristic blob temperature 
$T\approx$7~MK and the value of $log(EM)\approx$23~cm$^{-3}$) was determined. In later times than 17:25~UT, the $EM$ and~$T$ up growth caused saturated data (not presented) over whole studied region. Thus, the temperature $\geq $7~MK might be assumed, at least in the central part of the active region during the flare.

\begin{figure}[h]
\centering
\includegraphics[width=6cm]{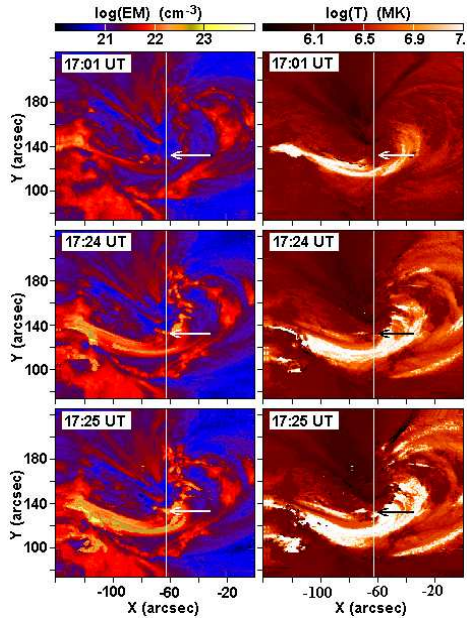}
\caption{Coronal AIA filtergrams at 17:01, 17:24, and 17:25~UT. Extrapolated parameters of emission measure~$EM$ and temperature~$T$ of the sunspot region. Vertical lines show the $X$~=~-62\arcsec~slit position. Arrows display the $Y$~=~132\arcsec position and the initial impulse location at 17:25 UT (bright blob).}
\label{fig08}
\end{figure}

\section{Discussion and conclusions}
We studied the isolated sunspot of positive magnetic field polarity observed near the solar disc centre during the X1.6-class flare with the SDO/AIA/HMI and IRIS instruments (Fig.~2). We examined the 16:20--18:20~UT interval. We found a magnetic field strength profile with a~hill shape and steep slopes ($(d)$, Fig.~7) along the 
$X$~=~$-$62\arcsec~slit where the highest magnetic field strength ($\approx$2\,200~G) was detected. This hill shape profile was also assumed for the plasma density profile and its stability can be maintained by the strong magnetic field. This plasma profile was filled by steady plasma flows, resulting from photospheric and chromospheric evaporation. The plasma flows and the tadpole wave continually propagated from the spot penumbra location ($Y~\approx$130\arcsec) towards the umbra light bridge 
($Y~\approx$150\arcsec), that is, in direction of the enhanced plasma density (Figs.~2,~4, and Movie~1). The tadpole wave appeared ($\approx$20~minutes after the plasma flow started) when the plasma flows created a~convenient environment in the waveguide, formed by gradients of the field-aligned speed of the plasma steady flows (Nakariakov \& Roberts 1995). 

The tadpole waves are impulsively generated and, therefore, we searched for a~possible initial perturbation. This was found at a~bright blob location, which was clearly visible in all observables at 17:25~UT (arrows, Figs.~1 and~7). Since some of the tadpole waves started sooner than 17:25~UT (Table~2) and a~real initial impulse lasts for a rather short time (few seconds), a~non-propagating entropy wave (Murawski at al. 2011) that occurred and stayed in situ of the initial perturbation may be a~more likely option for the observed blobs. The entropy wave peak only increases with time (our case) when the initial perturbation is caused in a~current sheet (M\'esz\'arosov\'a et al. 2014). This entropy wave (blob) could occur in the event of a~possible reconnection between a~loop and ribbon, for example. 

The vertical magnetic field lines above the spot ($(b)$, Fig.~7), connected the inner atmosphere with the solar corona, created a~robust magnetic field flux tube that made a~magnetically coupled atmosphere. It facilitated propagation of any changes that appeared at individual atmospheric layers. Any perturbations could be easily transported from the inner atmosphere without a~reflection before reaching the solar corona. Therefore, the common behaviour of flows and waves at all (E)UV observables (Figs.~2--4) was detected. This magnetic field tube was temporally affected by the passing tadpole wave, causing the forced reconnection of magnetic field lines (arrows, $(a)-(b)$, Fig.~7)) and a decrease in the tube magnetic field strength of $\approx$1\,200~G. Since they happened in the time of the flare X-rays maximum (GOES-15), these magnetic field perturbations could have caused the power of this flare. This is the first flare observation with an umbra magnetic field reconnection that was forced by an external (triggered in penumbra) tadpole wave of a~large period (e.g. $P_\mathrm{ch}$~=~1\,607~s in the solar corona). When the tadpole wave left its place, the spot magnetic field was restored to its original strength (17:45 UT, $(e)$, Fig.~7)). 

Usually, the plasma density in the photosphere is several orders of magnitude larger than that in the corona
and the magnetic field lines above a sunspot create a big canopy structure with a significantly lower magnetic field line density in a corona than in the photosphere. Therefore, the magnetic field strength B as well as the plasma density in the photosphere are significantly larger then in the corona. 

The situation is different when the magnetic flux tube is rather compact with a small canopy structure in the corona 
(17:20~UT, $(b)$, Fig.~7). The magnetic field line density and the plasma density are larger in the photosphere than in the corona, but they are not so different. This may be due to the fact that the sunspot is isolated; usually we observe 
a~group of sunspots in one active region. Also, the strongly magnetically coupled atmosphere, including the corona by the large value of the magnetic field strength B$\approx$2\,000~G, can play a~role. Thus, this situation is rather unusual and
was observed for the first time.

The $X$~=~$-$62\arcsec~slit played a~role in the longitudinal cylinder axis where the time series (Fig.~4) showed the highest values of intensity. 
We determined the tadpole waveguide in all atmospheric layers above the sunspot with the longitudinal length~$L$ and cross-section radius~$R$ in a~range of 11--18~Mm and 6--8~Mm, respectively (Table~2). 
A~dimensionless phase speed $a_\mathrm{ph}$ (Table~3) of individual tadpole waves were normalised on the waveguide internal Alv\'en speed (Nakariakov \& Roberts 1995): 
$a_\mathrm{ph} = \omega/k (v_\mathrm{A} + M_\mathrm{0}),$ where $M_\mathrm{0} = U/v_\mathrm{A}$ is the Mach number inside the cylinder and $U$ is the plasma flow speed (Table~1). This speed $a_\mathrm{ph}$ ranged from 1.02--1.79 and was highest in the inner atmosphere and decreased towards the solar corona (Table~3). It was caused by the plasma flows, which were faster in the inner atmosphere where they were generated (speed~$U$, Table~1, Fig.~3). Later, in the corona, the speed of plasma flows and the tadpole wave might be balanced. Similarly, the plasma flow speed and wave Alfv\'en speed (Table 1--2) is 15.38 \&~13.73~km~s$^{-1}$ in the photosphere and 10.99 \&~10.35 km~s$^{-1}$ in the corona, respectively. It seems that these speeds are also balanced with a~greater distance from the inner atmosphere where they were generated.

The typical Alfv\'en speed is several hundreds to several thousands of km~s$^{-1}$  in the solar corona. To our knowledge, the solar coronal waveguides studied thus far have been observed under different physical conditions (e.g. in coronal loops). The Alfv\'en speed is proportional to the waveguide internal magnetic field strength and inversely proportional to the square root of the plasma density. Since we do not know these values for the individual atmosphere layers above the sunspot, we used the maximal observed wave speed $v$ (Fig. 5) as an approximation for the $v_\mathrm{A}$.
Our low $v_\mathrm{A}$ values can be affected by the plasma flows propagating together with the tadpole waves
in the robust waveguide as well as by the strongly magnetically coupled solar atmosphere. MHD numerical simulations
with the propagating plasma flows and tadpole waves in a robust waveguide with the strong magnetic field strength might help us to understand them in a~better way.

The sausage wave periods (Table~2) can be seen to be much longer than typical values. But the sausage wave periods depend on the physical parameters of the waveguide (e.g. cross-section radius). Thus, a~robust wave with a~long period may be generated in a~robust waveguide.

The tadpole waves finished close the position of $Y \approx$150\arcsec~for most EUV observables because of the less suitable physical conditions for their propagation in the spot light bridge location (Fig.~7). In some cases (e.g. in 1600\AA~diagram), these waves were temporarily reduced around this location and later they continued (in non-reduced form) at places with higher $Y$~coordinates. Thus, the trapped tadpole waves of the inner atmosphere were able to overcome this location (see Movie~1 for 1600 and 1700\AA~observables) and it might be related to their higher Alfv\'en speed $v_\mathrm{A}$ (Table~2).

The regular, leaky, tunnelled tadpole waves can often occur in active regions since the flare impulsive phase provides numerous possible impulses that can trigger these types of waves. Moreover, the solar corona usually contains many enhanced plasma density structures that can form waveguides. The trapped tadpole wave can act as a~moving source of the leaky waves and their impulsively deposited energy is released outside original waveguide (Nistic\`{o} at al. 2014). Particularly, the leaky tadpole waves with the periods of $P \approx$1600--2100~s could play a~role in the initial impulses to generate the next generation of the trapped tadpole waves outside the original waveguide. If this process can be repeated, the trapped, leaky, and tunnelled waves (Fig.~6) could be dissipated throughout the active region. 

Sometimes, the trapped wavelet tadpole signatures exhibited additional components (fins), which are associated with their heads (M\'esz\'arosov\'a et al. 2014; Shestov et al. 2015). For example, some of the fins (Shestov et al. 2015) resemble the 1700 and 335\AA~tadpole wave fins in Fig.~4. Examples of changes in the fins morphology are displayed in Movie~1 for AIA 171, 304, 335, 1600, and 1700\AA~observables. These fins may reflect physical changes in the individual waveguide plasma conditions, which are not well understand yet. 

The dispersive nature of the tadpole waves with their easy ability to generate the leaky and other modes propagating outside the original waveguide and magnetic field flux tubes connecting the individual atmospheric layers can distribute the photospheric and chromospheric magnetic field energy across the active region.
This mechanism can contribute to the coronal energy balance and to our knowledge as to how the coronal heating is maintained. For a deeper understanding on the processes of flare magnetic energy distribution over the solar corona, the parametrised MHD numerical simulations are needed to know more about the connection between individual tadpole wave modes and between the plasma flows and tadpole waves propagating in the same place and time. Also new observations (e.g. DKIST, Solar Orbiter) can deepen our knowledge of the coronal magnetic fields, the magnetically coupled solar atmosphere, and structures of enhanced plasma density (waveguides). 

\begin{acknowledgements}
We acknowledge the Bilateral Mobility Projects SAV-16-03 and SAV-18-01 between the Academy of Sciences of the Czech Republic and the Slovak Academy of Sciences. H.M.
acknowledges support by the project RVO: 67985815. P.G. acknowledges the financial support by the Science Grant Agency project VEGA 2/0048/20. We thank Dr. J. Ryb\'{a}k for
his help on the SDO/AIA and IRIS data processing and the artificial slit numerical code. This article was created by the realisation of the project ITMS No. 26220120029, based on the supporting operational Research and development program financed from the European Regional Development Fund. We thank the SDO and IRIS teams for obtaining the data. The SDO is a~mission for NASA Living With a~Star (LWS) Program. This research has made use of NASA Astrophysics Data System. IRIS is a~NASA small explorer mission developed and operated by LMSAL with mission operations executed at NASA Ames Research centre and major contributions to downlink communications funded by ESA and the Norwegian Space Centre. The wavelet analysis was performed using the software based on tools provided by C.~Torrence and G. P. Compo at \texttt{http://paos.colorado.edu/research/wavelets}. The method tutorial and performance tests using the Vertical Current Approximation Non-linear Force-free Field method (VCA-NLFFF) for non-potential field modelling of coronal loop and an automated tracing of coronal loops are available at $http://www.lmsal.com/aschwand/software/nlfff/nlfff\_tutorial1.html$. The SDO/AIA emission measure and temperature method is described in the available tutorial at $http://www.lmsal.com/aschwand/software/aia/aia\_dem.html$.
\end{acknowledgements}

\begin{appendix}
\section{WAvelet SEparation Method (WASEM)}
\begin{figure*}[ht]
\centering
\includegraphics[width=18cm]{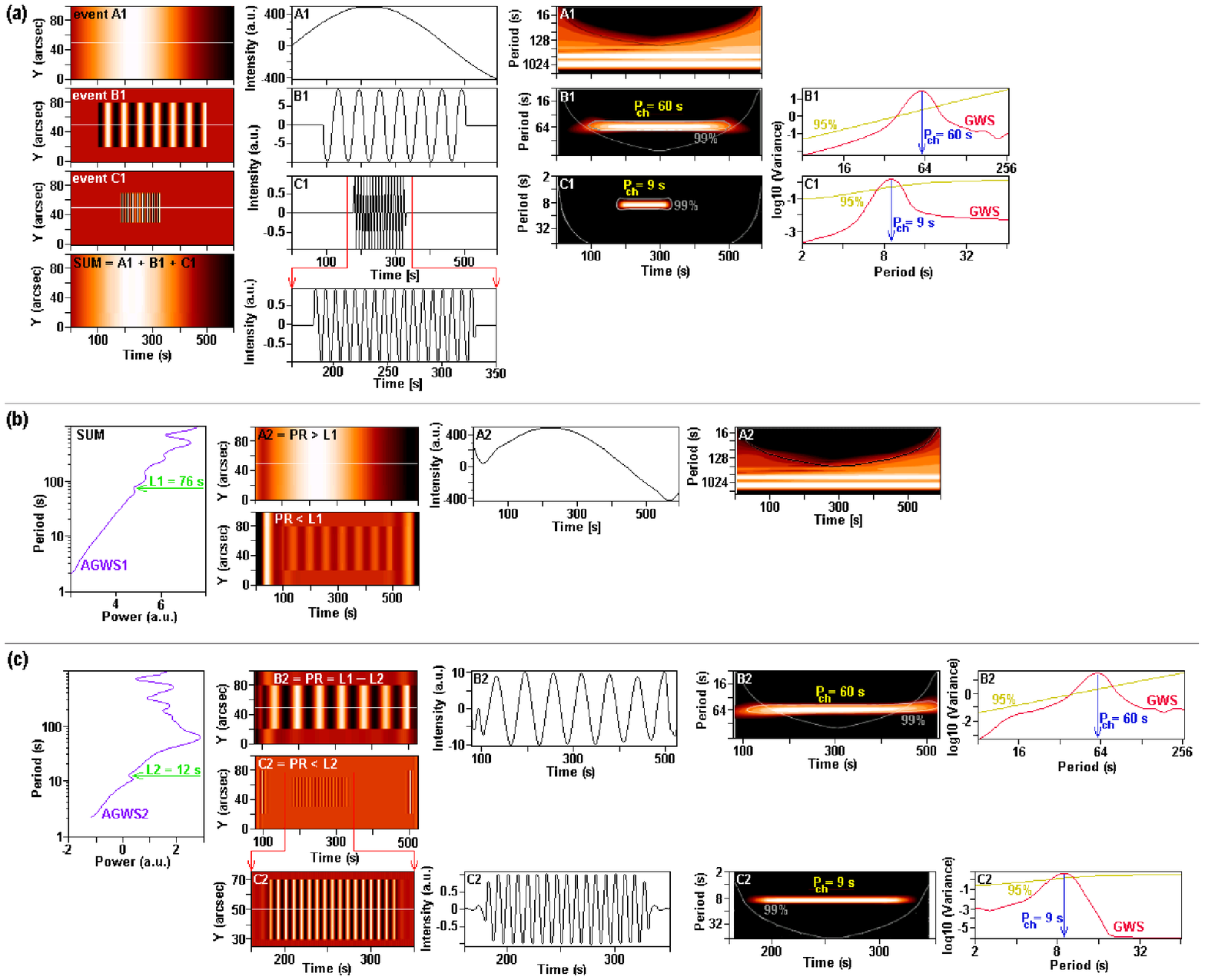}
\caption{WASEM separation of artificial space-time diagram $SUM$. Vertical rows show the event with a~selected time series at $Y$~=~50\arcsec, the wavelet power spectrum, and the global wavelet spectrum GWS with characteristic periods $P_\mathrm{ch}$ above the global 95\% significance level curve. $PR$ is the selected period range.
Part~$a$: Emergence of $SUM = A1 + B1 + C1$ events.  
Part~$b$: Event~$A1$ separation according to the averaged global wavelet spectrum AGWS1 with $L1$~limit. 
Separated space-time diagrams $A2=PR>L1$ and $PR<L1$. Selected time series of $A2$ with corresponding wavelet power spectrum. 
Part~$c$: Events $B1-C1$ separation. AGWS2 with $L2$ limit. Separated space-time diagrams $B2=PR=L1-L2$ and $C2=PR<L2$. Selected time series of $B2$ and $C2$ with corresponding 
wavelet power spectra.}
\label{fig09}
\end{figure*}

\begin{figure*}[ht]
\centering
\includegraphics[width=14cm]{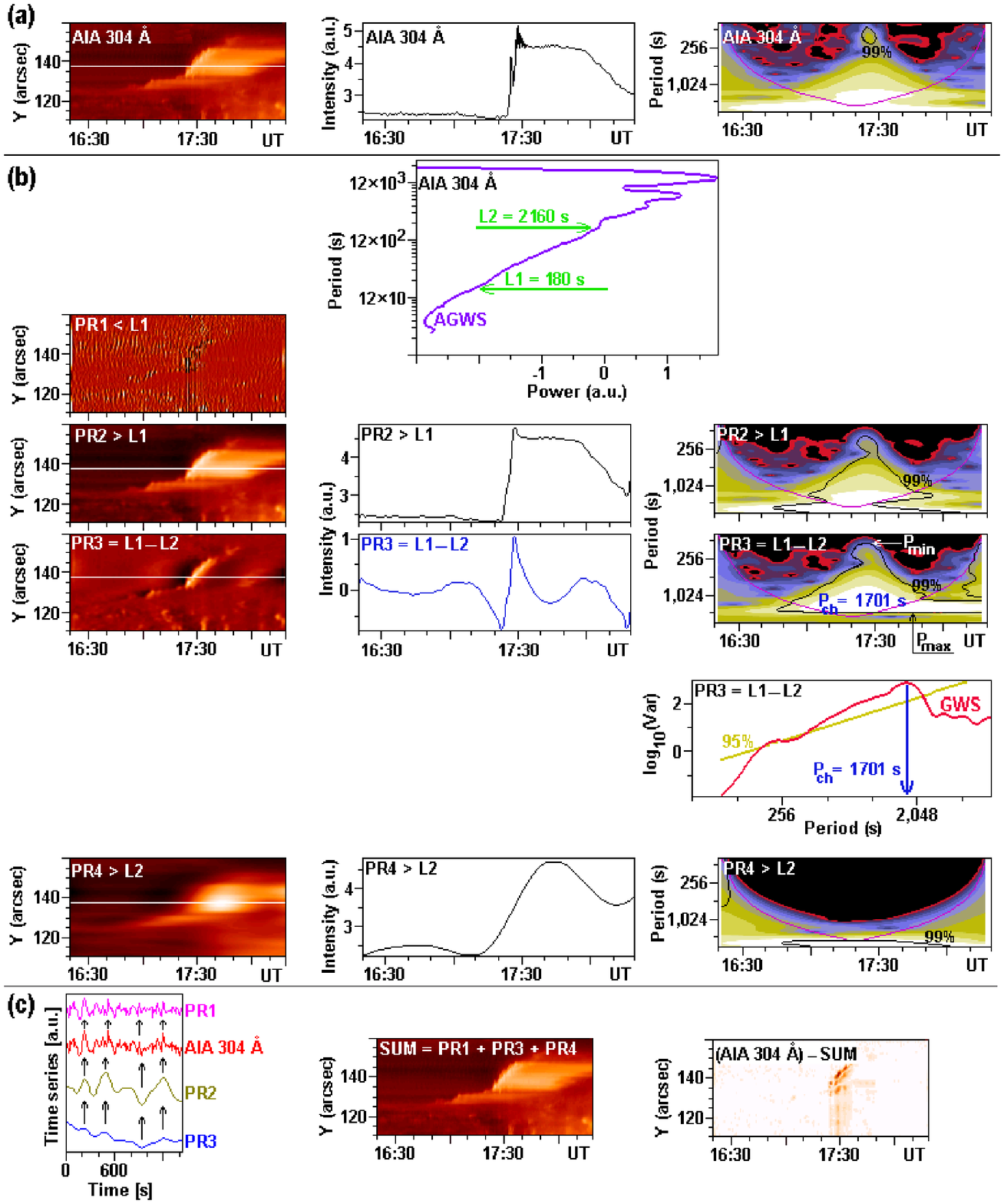}
\caption{WASEM separation of the observed space-time diagram SDO/AIA 304\AA ($(a)$) with selected time series at $Y$~=~138\arcsec and wavelet power spectrum. 
Part~$(b)$: Averaged global wavelet spectrum AGWS with $L1-L2$~limits. Separated space-time diagrams according to period range $PR1-PR4$ ($PR3$ and $PR4$ with the tadpole wave 
and the plasma flow propagation, respectively). Global wavelet spectrum (GWS) determined a~characteristic period $P_\mathrm{ch}$ above the global 95\% significance level curve. WASEM analysis validity tests (part~$(c)$:).}
\label{fig010}
\end{figure*}

The WAvelet SEparation Method (WASEM), based on the wavelet analysis techniques, is useful for the analysis of observational data maps consisting of two or more individual (quasi-)periodic physical phenomena. The data map can contain (a)~non-periodic emission(s). Thus, WASEM is suitable when an original data map (e.g. imaging data, space-time diagram, dynamic spectrum) consists of a~mixture of different structures that are observed at the same location and frequencies during the same time interval. In such a~case, weaker but interesting physical phenomenon may coincide with a~strong emission. Usually, this strong component is well detectable, while the weak one remains hidden. This type of~situation is characteristic of complex (e.g. radio) spectra that are observed during solar flares, which can consist of different types of bursts, plasma jets, or MHD waves and oscillations (M\'esz\'arosov\'a et al. 2016). When we want to separate individual temporal or spatial components from one another, we can also use WASEM to learn about the hidden structures, if they exist that is. The separated physical phenomena can be easily analysed with a more precise determination of their parameters. This method can also be used for instrumental interference filtering. Formerly, a~similar method (M\'{e}sz\'{a}rosov\'{a} at al. 2011) was used for a~separation of different bursts in radio dynamic spectra, according to their different frequency drifts.

In this study, we used WASEM to reveal steady plasma flows and tadpole waves hidden in the time series of the space-time diagrams that were computed from the original SDO/AIA and IRIS imaging data. Particularly, the tadpole waves were not directly visible since they propagated inside a~waveguide filled with dense and hot plasma, generating their own strong (E)UV emission. We used the Morlet wavelet power spectrum (Torrence \& Compo 1998) for our analysis. The confidence level relative to red noise was taken into account and the solid contour around the tadpole patterns shows the 99\% confidence level (Fig.~4). Thus, we only studied the most dominant characteristic wavelet signatures. The wavelet spectra were plotted with the lighter areas indicating the greater power. The region under magenta curves in the wavelet power spectra (Fig.~4) belongs to the cone of influence (COI) where edge effects can become important as a result of dealing with finite-length time series. Wavelet parameter $\omega$ was set to~6, that is, we preferred to specify an exact period rather than an exact time interval. Newly reconstructed space-time diagrams with the separated individual plasma flows or tadpole waves were computed with the~help of the inverse wavelet analysis (Torrence \& Compo 1998) for individual period ranges~$PR$. The positive and negative parts of the amplitudes in these new space-time diagrams (in relation to their mean values) are given in white and black, respectively. The WASEM analysis was applied to all of the original SDO/AIA and IRIS space-time diagrams (example in Fig.~A.2 $(a)$) computed at individual $X$-coordinates under study. The same WASEM wavelet parameters were used in this whole study to get comparable results together with each other. 

\subsection{WASEM in numerical simulations}
The WASEM possibilities are exhibited using numerical simulation (Fig.~A.1). The part~$(a)$ describes the emergence of the artificial space-time diagram $SUM$, consisting of three different events $A1-C1$. The parts $(b)-(c)$ display how to divide the diagram $SUM$ into separated ones ($A2-C2$). Finally, we compare the separated results ($A2-C2$) with the original events ($A1-C1$). 
The individual rows in Fig.~A.1 consist of the space-time diagram and the selected time series at $Y$~=~50\arcsec~with its wavelet power spectrum. Moreover, the $AGWS$ curve shows the averaged global wavelet spectrum computed for all time series of the space-time diagram. It shows the most dominant periods found in the entire space-time diagram. This curve informs us about (quasi-)periodical processes in our data set. The $GWS$ curve (Torrence \& Compo 1998) displays a~global wavelet spectrum computed for the selected time series to determine a~characteristic period $P_\mathrm{ch}$ above a~wavelet significance level of $>95\%$. 

We generated three artificial space-time diagrams (part~$(a)$, Fig.~A.1) where the global time interval~is~0--600~s, the spatial range is~$Y$~=~0--100\arcsec, the temporal cadence is~1~s, and the spatial resolution is of 1\arcsec. These diagrams contain the following: (i)~event~$A1$: strong continuum emission with an intensity of 500~a.u. (arbitrary units) and a~period of 900~s; 
(ii)~event~$B1$: a group of pulsations with a~period of 60~s, the emission intensity of 10~a.u., and observed time interval~of~100--500~s at coordinates $Y$~=~20--80\arcsec; and (iii)~event~$C1$: a small group of weak pulsations with a~period of 9~s, an emission intensity of 1 a.u., and an observed time interval~of~180--330~s at coordinates $Y$~=~30--70\arcsec. The selected time series of the $A1-C1$ events are marked by a~vertical line at $Y$~=~50\arcsec~in their space-time diagrams. Their wavelet power spectra show the characteristic periodicity. The global wavelet spectra $GWS$ enable us to determine a~characteristic period $P_\mathrm{ch}$ of 60 and 9~s for the events $B1$ and $C1$, respectively. The resulting space-time diagram $SUM$ (part~$(a)$) is a~summary of these events $A1+B1+C1$, where the events $B1$ and $C1$ are invisible and hidden below event~$A1$. This is a~common occurrence in observed data.

Then we applied WASEM to the input data $SUM$ to exhibit the possibility of detecting and separating the hidden $B1$ and $C1$ events. The global wavelet spectra $GWSs$ for all of the individual $SUM$ time series were computed and integrated to get an averaged global wavelet spectrum $AGWS1$ (part~$(b)$). This curve displays the most dominant periods (curve peaks), which were found in the space-time diagram $SUM$. We selected a~limit $L1$ of 76~s for the data separation. The new separated space-time diagrams were computed for the period ranges $A2=PR>L1$ and $PR<L1$. Practically, we selected only time series of the individual period range $PR$ from the diagram $SUM$ to put them into the new separated diagram. Thus, for example, new diagram $A2$ only consists of the time series with $PR>L1$. Now the strong continuum is separated in the new $A2$ space-time diagram. To prove this, we selected the time series marked by a~vertical line (as an example) to compute its $A2$~wavelet power spectrum. We gained a~good separation since the event $A2=A1,$ except for the left and right narrow margins of the separated $A2$ time series, because of an effect of finite time series length (Torrence \& Compo 1998).

Now we can see some pulsations in the space-time diagram $PR<L1$ (rest of separation), which became visible just after filtering $A2$ data out. Therefore, we repeated the entire process to make sure there were not any other weaker structures that were hidden. The global wavelet spectra (GWSs) for all of individual time series of the $PR<L1$ diagram were computed and integrated to obtain the averaged global wavelet spectrum AGWS2 (part~$(c)$). Then we selected a~limit $L2$~=~12~s for data separation with the period ranges $B2=PR=L1-L2$ and $C2=PR<L2$. 
The new separated space-time diagrams $B2$ and $C2$ were computed. The selected time series (marked by a~vertical line) with their wavelet power spectra are displayed (part $(c)$). 
We gained good separations where the event $B2=B1$ and $C2=C1$ and we determined characteristic periods 
$P_\mathrm{ch}$~=~60~s (for $B2$) and $P_\mathrm{ch}$~=~9~s (for $C2$). Thus, the WASEM separation of all events ($A1, B1$, and $C1$) from one another was successful. 

\subsection{WASEM in the observed data}
This WASEM analysis was applied to the space-time diagrams of all (E)UV observables in the same way. This is demonstrated for the original SDO/AIA~304\AA~space-time diagram 
($(a)$, Fig.~A.2), where the selected time series at $Y$ = 138\arcsec~is marked by a~vertical line. The wavelet power spectrum of this series was not very informative; there was only a~small area with 99\% significance. No valid decision regarding the presence of, for example, waves and oscillations could be made on the basis of this wavelet pattern. In this case, WASEM can be useful.

Therefore, the global wavelet spectra ($GWS$) for all individual time series presented in the original space-time diagram (AIA~304\AA) were computed and integrated to get 
the averaged global wavelet spectrum ($AGWS$, $(b)$).  Since the tadpole wave train consists of short-, middle-, and long-period components, the $AGWS$ curve exhibited a~straight part with no dominant peaks. This type of straight part may be the first signature of the tadpole wave presence in data under study. This $AGWS$ part was marked by arrows for selected period range limits $L1$ and $L2$ of 180 and 2\,160~s, respectively. They were used for the separations in these ranges: $PR1<L1$, $PR2>L1$, $PR3=L1-L2$ (i.e. period range from $L1$ up to $L2$), and $PR4>L2$. Newly reconstructed space-time diagrams ($(b)$) were computed. Their time series at $Y$~=~138\arcsec~were selected and their corresponding wavelet power spectra were computed. The new $PR<L1$ space-time diagram consisted of possible instrument interferences. The $PR>L1$ diagram showed the original diagram without these interferences. Its wavelet power spectrum of the same selected time series was more informative than the original one. It displayed a~possible tadpole pattern, except for the long-period components that still remain unclear because of the intensive (E)UV emission continuum in the original space-time diagram.

The new $PR=L1-L2$ space-time diagram is cleaned from the interferences as well as from the intensive emission continuum. Its wavelet power spectrum exhibited a~characteristic cylindrical tadpole wave pattern (Shestov et al 2015). We determined the characteristic tadpole wave period $P_\mathrm{ch}$ of 1\,701~s as a~period of $GWS$ peak above the global 95\% significance level ($(b)$, Fig.~A.2). The maximal and minimal periods ($P_\mathrm{max}$ and $P_\mathrm{min}$) were determined according to the tadpole pattern in their wavelet power spectrum (Table~2). Similar tadpole wave patterns were found for all of the (E)UV observables (Fig.~4, Movie~1).

The $PR>L2$ space-time diagram showed just the intensive drifting (E)UV emission. The selected time series displayed a background continuum and its wavelet power spectrum did not have any relevant periodicities. These separated space-time diagrams were used for an analysis of the (E)UV steady plasma flows (Fig.~3).

The WASEM validity tests are shown in this paper ($(c)$, Fig.~A.2). (i)~A~new space-time diagram consists of reconstructed time series with peaks just existing in the original series (i.e. no important artificial peaks are generated numerically). A~comparison of the individual peaks of the original AIA~304\AA~space-time diagram (in red) and of the reconstructed $PR1$--$PR3$ (other colours) is shown for the selected time series at $Y$~=~138\arcsec~(left panel, $(c)$). To display the individual peaks, we selected a~shorter time interval of 16:28--16:52~UT (duration of 1\,440~s) and corresponding peaks are marked by arrows. (ii)~A~sum of all separated space-time diagrams, that is, $SUM=PR1+PR3+PR4$ (middle panel, $(c)$) should be equal to the original diagram (AIA~304\AA). The difference between these two diagrams (right panel, $(c)$) exhibited negligible WASEM numerical interferences. 

WASEM has two period constraints. (i) Firstly, the smallest detectable period $P_\mathrm{s}$ is a constraint since one quarter of a~pulse (2$\pi$/4) needs this minimal number of temporal samples for its definition. The rest of the pulse is reconstructed with knowledge of the first quarter (Torrence \& Compo 1998). We used at least twice the data time cadence (e.g. 24.79~s instead of 12~s of the AIA 304\AA~observable) for the smallest possible period $P_\mathrm{s}$ during the separation process. Therefore, the periods $< P_\mathrm{s}$ cannot be contained in the $SUM$ diagram.
(ii) Secondly, the highest detectable period $P_\mathrm{h}$ is the second constraint and it equals four times the total time interval (Torrence \& Compo 1998). We used the two-hour time interval under study. Its quadruple is eight hours = 288\,000~s. With the data cadence of 12~s, the highest detectable period $P_\mathrm{h}$ is 24\,000~s (= 288\,000/12). These types of periods are always non-real since they are derived from the whole time series length. Thus, the lowest $P_\mathrm{s}$ and highest $P_\mathrm{h}$ possible periods that can be detected depend on the data temporal cadence and the time series length, respectively. Then the wavelet analysis can estimate the most likely period in the range of $P_\mathrm{s}$--$P_\mathrm{h}$, that is to say, these periods have a~probabilistic character, which depends on the selected wavelet parameters (e.g. red noise, selected significance). WASEM separates individual physical phenomena according to their different periodicities found in the data set. In such a~case, when two physical phenomena have the close periodicity features, WASEM cannot recognise them from one another successfully. There is no universal rule regarding how to determine convenient $AGWS$ limits~$L$ because of the different physical properties of individual subjects under separation. Different period ranges $PR$ between the limits~$L$ should be tried to reveal reasonable  hidden structures. Knowledge of the physical phenomenon under study can help do this. For example, the characteristic wavelet pattern of the tadpole wave is well-known. Then, we need to check  the smallest tadpole period ($P_\mathrm{min}$) carefully to see the whole tadpole head with its possible fins as well as the highest tadpole period ($P_\mathrm{max}$), which might reveal the entire bottom part of the tadpole body. This is important for the characteristic ($P_\mathrm{ch}$) and maximal ($P_\mathrm{max}$) period determination. 

\section{VCA-NLFFF method and used parameters}
\begin{figure}[h]
\centering
\includegraphics[width=8cm]{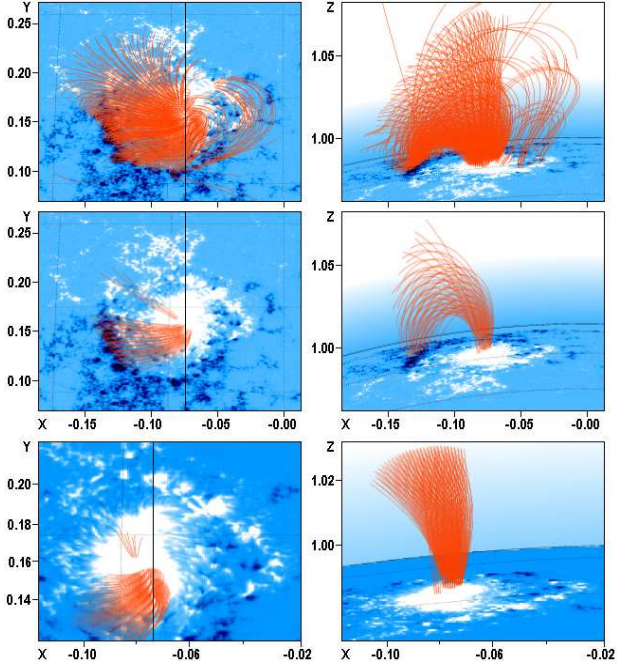}
\caption{Non-linear force-free magnetic field lines (VCA-NLFFF) above the sunspot at 16:50~UT. The blue image background represents the SDO/HMI magnetogram with the sunspot (in white) and magnetic field lines computed in the line-of-sight plane (left column) and rotated to the north of 90 degrees (right column). Vertical black solid lines show the $X$~=~-62\arcsec~slit position. The magnetic field lines (red curves) were computed with parameters: $ngrid$~=~120, 150, and 150 \& $bmin$~=~150, 180, and 180 in the upper, middle, and bottom panels, respectively.}
\label{fig11}
\end{figure}
We used The Vertical-Current Approximation Non-linear Force-Free Field code (VCA-NLFFF; Aschwanden 2016) to make the evolution of non-linear force-free magnetic field lines clear above the observed sunspot. For the magnetic field line representation, the rectangular grid of footpoints (parameter $ngrid$) was chosen with a~threshold set by the minimum magnetic field strength (parameter $bmin$) of the NLFFF\_DISP2 procedure. Time sunspot snaps of 16:50~UT are displayed in Fig.~B.1 
where the blue image background represents the $B_\mathrm{z}$~SDO/HMI magnetogram with the sunspot (in white). The best-fit magnetic field lines (red curves) were computed from two different aspect angles: 
(i) the line-of-sight $X$-$Y$ plane of the sun surface (left column) and (ii) the same figure rotated to the north of 90 degrees  (right column) where the vertical $Z$~coordinate means the altitude above the solar surface. Thus, these images (right column) exhibited the magnetic field connection between the photosphere and the solar corona (i.e. magnetic flux tube above the spot). In all cases, the coordinates are expressed in solar radii. The vertical solid lines (left column) show the slit position of $X$~=~-62\arcsec.
We used these values for the other VCA-NLFFF parameters as follows: the maximum altitude $hmax = 0.2 R_\odot$, the number of magnetic sources $nmag\_p$~=~100, the minimum number of iterations $nitmin$~=~40, and the maximum number of iterations $nitmin$~=~100.

The magnetic field lines (upper panels, Fig.~B.1) for the sunspot region ($NOAA$~=~'12158', $helpos$~=~'N17E05', $FOV$~=~0.2) included the pores (in dark blue) of a~negative magnetic field polarity. The magnetic field lines were computed with footpoints 
in a~120~$\times$~120 rectangular grid ($ngrid$~=~120), and with field strengths of $|B_\mathrm{z}| >$~150~G ($bmin$~=~150; expressed in Gauss). These panels exhibited many magnetic field lines and, therefore, any possible magnetic field perturbation cannot be visible. A~better representation (middle panels) also showed the sunspot region with the negative pores, but the magnetic field lines were computed with the footpoints in a~150~$\times$~150 rectangular grid ($ngrid$~=~150) with field strengths of $|B_\mathrm{z}|>$~180~G ($bmin$~=~180). Thus, only the magnetic field lines between the slit position of $X$~=~-62\arcsec~and the negative pores are displayed. Also, in this case, a~possible magnetic field perturbation cannot be clearly visible.

The magnetic field lines (bottom panels, Fig.~B.1) were calculated with the same parameters ($ngrid$~=~150 \& $bmin$~=~180) for the same part of the active region, but without the negative pores ($NOAA$~=~'12158', $helpos$~=~'N17E04', $FOV$~=~0.1). In this case, more magnetic field lines were exhibited. The left panel displayed a~highly twisted field around the position of $X$~=~-62\arcsec and the right panel showed the characteristic magnetic field canopy above the spot. 
This version of the magnetic field line representation was used ($(a)-(b)$, Fig.~7) to reveal perturbations of the magnetic field lines during the time interval of the tadpole wave propagation.

\end{appendix}

\end{document}